\documentclass[fleqn,12pt,twoside]{article}
\usepackage[headings]{espcrc1}

\readRCS
$Id: espcrc1.tex,v 1.2 2004/02/24 11:22:11 spepping Exp $
\usepackage{graphicx}
\usepackage[figuresright]{rotating}

\newcommand{\AmS}{{\protect\the\textfont2
  A\kern-.1667em\lower.5ex\hbox{M}\kern-.125emS}}

\title{
Intriguing Trends in Nuclear Physics Articles Authorship
}

\author{B.\ Pritychenko\address[BNLab]{National Nuclear Data Center, Brookhaven National Laboratory, Upton, NY 11973-5000, U.S.A.}
}
       
\runtitle{ 
Intriguing Trends ...
}
\runauthor{B. Pritychenko}

\begin{document}

\maketitle

\begin{abstract}
The increase in authorship of nuclear physics publications has been investigated using the large statistical samples. 
This has been accomplished with nuclear data mining of nuclear science references (NSR) and experimental nuclear reaction (EXFOR) databases. 
The results of this study will be discussed and conclusions will be given.
\end{abstract}

\section{Introduction}
The authorship in physics publications is a very interesting topic that has been extensively discussed in 
Physics Today in recent years \cite{12Wy,12Au}. The journal readers disclosed many interesting observations that, 
unfortunately, are often based on rather limited statistics.   
``The ongoing obsession with citation count as the marker of achievement" \cite{12Au},  as well as pressures to ``publish or perish" are often 
cited as reasons for increases of authors number per paper \cite{84Sa}. Obviously, these reasons contribute to the presently-observed author 
list increase,  and these findings are supported by R. Heras \cite{13He}. 
At the same time, research authorship and its evolution over the years are complex phenomena that require extensive studies of scientific publications 
and broad discussions. To extend the scope of the above-mentioned studies, I will  investigate the authorship of  nuclear physics articles 
using the statistically-significant data samples extracted from the modern low-, and intermediate-energy nuclear physics databases  maintained 
by the National Nuclear Data Center (NNDC) \cite{12Pri}.

\section{Nuclear Physics Databases}

Collection and storage of nuclear bibliography materials is a foundation for nuclear data compilations and evaluations.  These extensive collections were assembled by nuclear data compilers over the past 50-60 years 
and represent a treasure trove for the scientists, who search these databases for nuclear  physics publications and relevant data.  In this work,  
I would consider the NSR ({\it http://www.nndc.bnl.gov/nsr})  and EXFOR ({\it http://www.iaea.org/exfor}) databases \cite{11Pri,14Ot}  
that contain records of more than 215000 publications, 92600 individual authors, and 20000 cross nuclear reaction measurements.     
These databases provide a complete coverage of nuclear physics publications, and are relatively-clean from the high-energy physics papers, where the authorship rules  are very different \cite{12Wy,14Aa}. 

Those maintained by the NNDC, NSR database \cite{11Pri} is a prime source of nuclear bibliography, and an excellent resource for investigation of the  
evolution of authorship from 1896 to the present days.  The EXFOR database \cite{14Ot} is more diverse 
than NSR because, besides, bibliographical information and keywords, it includes information on authors affiliations, experimental facilities, and results of the actual 
measurements.  It groups separate nuclear reaction publications that represent a single experiment into  EXFOR entry. 
Therefore,  in addition to the bibliographical information  since 1935, EXFOR also contains data on the number,  and geographical distribution of distinct 
experiments.    Historically, EXFOR compilations have been conducted internationally under the auspices of the International Atomic Energy Agency 
 in four major geographical areas:  area 1 (US and Canada), area 2 (Europe), area 3 (Asia, Australia, Africa and Latin America), 
and area 4 (Soviet Union/Russian Federation).

In order to extract the authorship, affiliation and experimental facilities information, NSR and EXFOR database contents, as of September 2014,  
have been examined with nuclear data mining techniques. A specialized-Java code with embedded SQL queries has been written to access 
the database contents, retrieve and process the metadata. The results of this study are presented below.  

\section{Analysis of Database Contents}

\begin{figure}[!htb]
\includegraphics[height=12cm, angle=0]{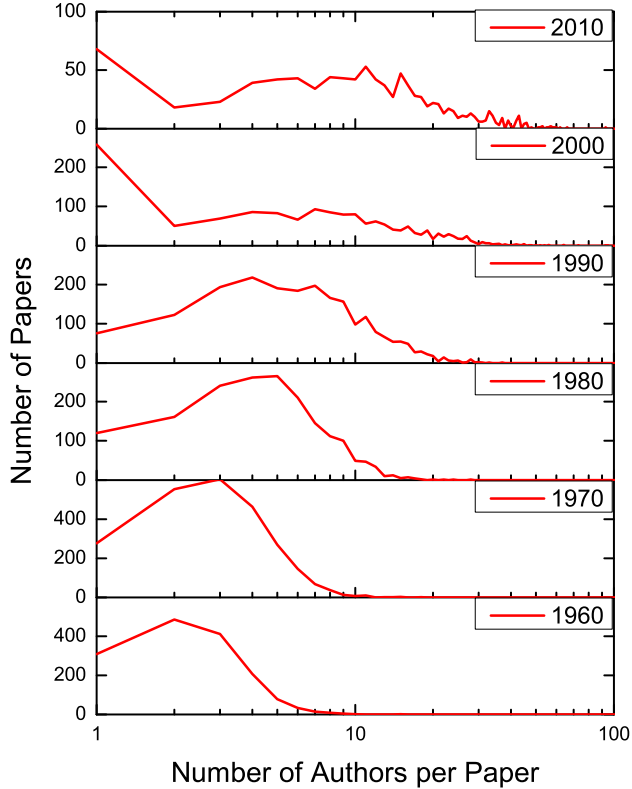}
\caption{Evolution of number of authors per paper in nuclear physics publications in 1960-2010. Data were taken from the NSR database \cite{11Pri} .}
\label{fig1}
\end{figure}

Following the pattern described by P.J. Wyatt \cite{12Wy}, author would analyze the evolution of authorship using the NSR bibliography data. 
Fig. \ref{fig1} shows an analysis of NSR content for several time periods.  In the early years, these data experience fluctuations are due to insufficient number of publications, 
and become very reliable in the 50ies. The Figure indicates that initially  
the majority of articles had from 1 to 3 authors, and nowadays it is a much larger number. These results are consistent with the 
many previously-reported findings \cite{12Wy,12Au,84Sa}. Furthermore, a detailed analysis of the older experimental 
papers also indicates that authorship rules in the past were very different, and names of many significant contributors 
were often kept in the  acknowledgements.  The analysis of acknowledgements shows that the following single contributions to 
experimental papers were not considered  sufficient for authorship in the past 
\begin{itemize}
\item Problem formulation and overall project supervision
\item Participating in data taking
\item Data analysis work, including code writing and running theoretical calculations 
\item Target preparation and help with detectors and accelerators
\end{itemize}
Authors were usually involved in multiple activities for the entire duration of the project.

\begin{figure}[!htb]
\includegraphics[height=12cm, angle=0]{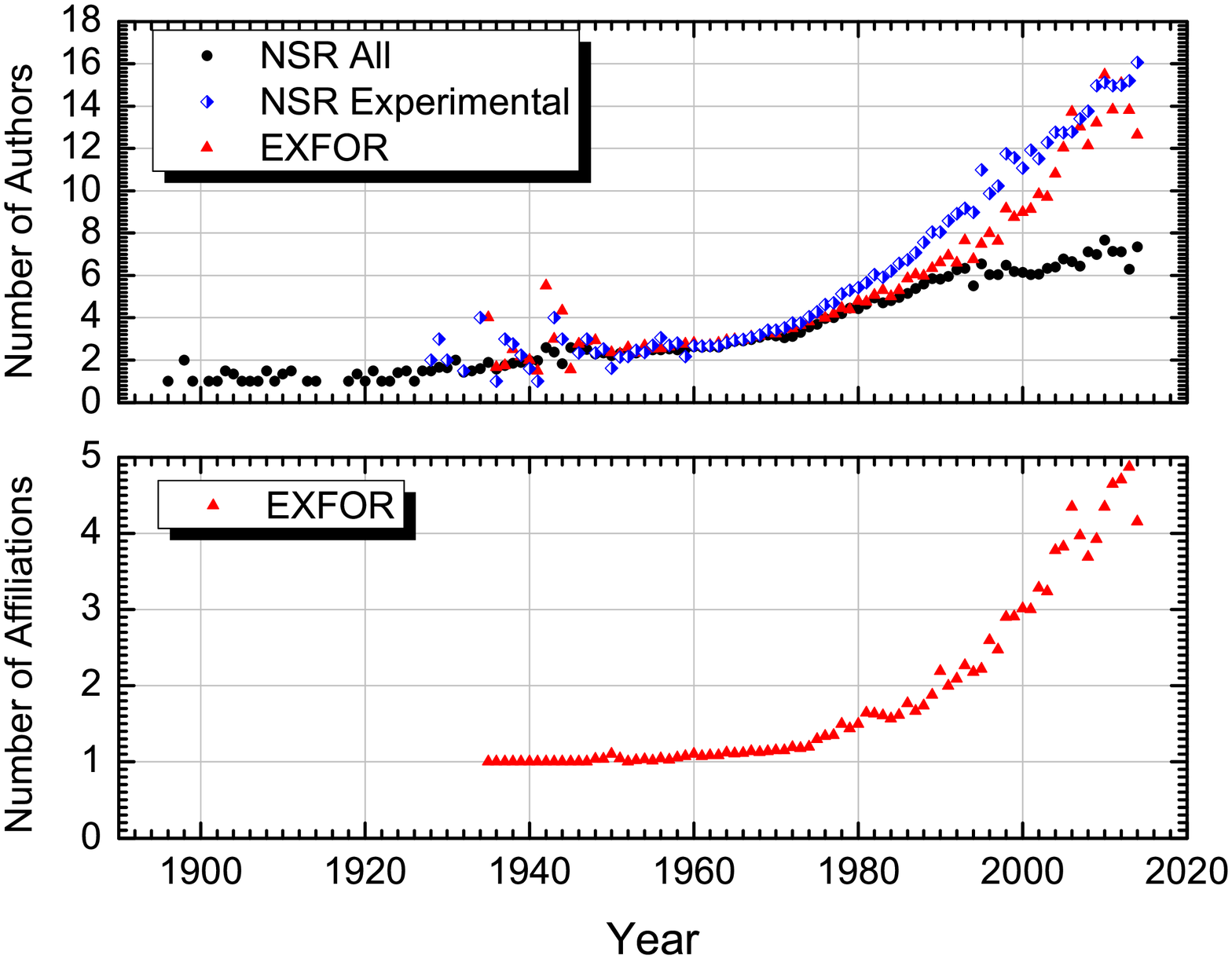}
\caption{Number of authors per article for NSR, NSR experimental, and EXFOR data collections is shown in the upper panel. The lower panel shows the distribution of affiliations documented in the EXFOR database \cite{14Ot} .}
\label{fig2}
\end{figure}
To extend the scope of the previous research, I would add affiliation data, and combine the bibliographical information in 
both databases with the actual number of research collaborations, which are  shown in the upper  and lower panels of Fig. \ref{fig2}, respectively. 
Fig. \ref{fig2} compliments the previous findings that the number of authors  per a single paper has increased dramatically in the last 50 years \cite{12Wy,12Au,84Sa}.  
Applying the  Moore's law  methodology \cite{14Mo},  one can deduce that number of authors doubles every 20-30 years. In some cases, EXFOR author numbers are lower than 
NSR because EXFOR has a substantial number of missing nuclear radioactive beam publications (Z$>$6). This explains the deviations of EXFOR authorship  
from NSR experimental papers in 1980-2000.  
The Figure also shows that authorship of experimental papers grows faster than authorship in general, and increases dramatically in 80ies and 90ies. 
The increase coincides in time with the end of ``Cold War" and Information Age revolution that has produced Email and Web collaboration tools.  
Shown in the lower panel of the Figure \ref{fig2}, number of affiliations for experimental papers initially has grown very slow, and 
reached two and three institutions per paper by 1990 and 2000, respectively.  These data provide a strong proof  for  the collaborative 
nature of research in recent years.

\begin{figure}[!htb]
\includegraphics[height=12cm, angle=0]{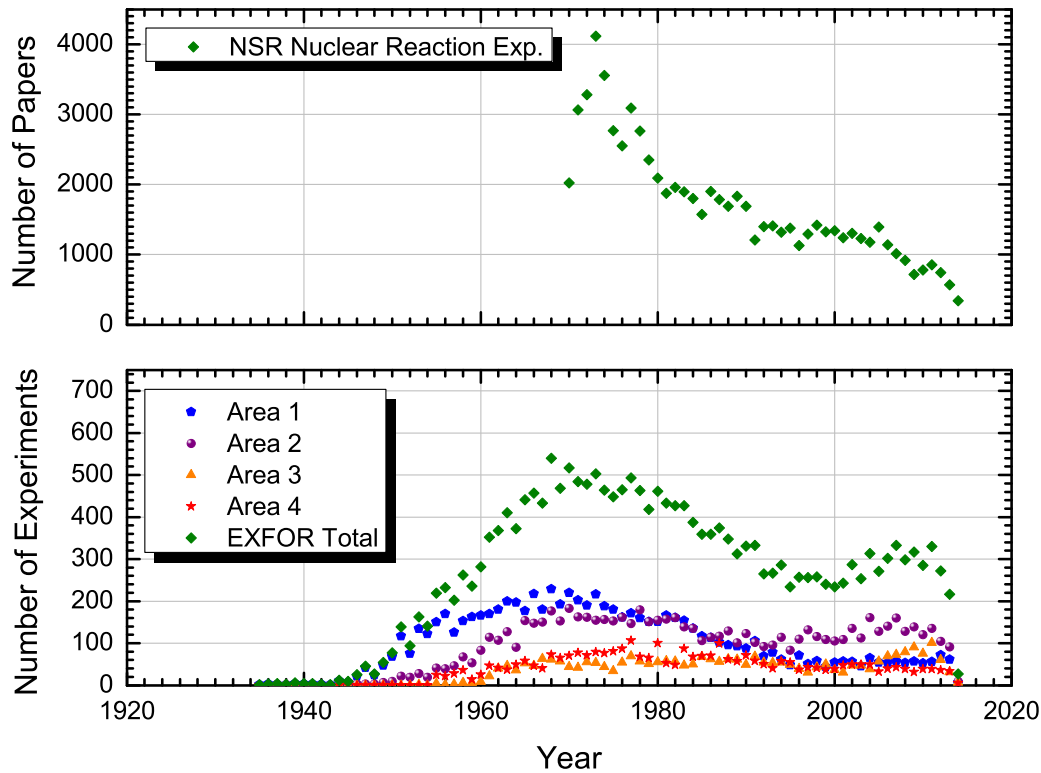}
\caption{Number of experimental nuclear reaction papers in NSR database \cite{11Pri}  is shown in the upper panel. The lower panel shows the distribution of nuclear reaction  measurements documented in the EXFOR database \cite{14Ot} .}
\label{fig3}
\end{figure}

Next, I would analyze an annual number of published nuclear reaction papers which is shown in the upper panel of Fig. \ref{fig3}. 
Those extracted from the NSR database numbers demonstrate decreasing or ``deflation" trend. This rather unexpected result requires 
an additional analysis of research facilities and the actual number of cross section measurements as  shown in the lower panel of Fig. \ref{fig3}. 
The data  indicate that decrease of experimental papers coincides with the reduced number  of measurements.  
The overall reduction of nuclear reaction  measurements worldwide in the last 20-30 years reflects a reduced number of operating facilities; when several North American accelerators (EXFOR area 1) were either shutdown 
or converted for different needs (Chalk River, ORELA, Yale, Indiana, ...). 
At the same time, the remaining operations have evolved into highly-specialized  user facilities. 
These trends are consistent with the M.  Thoennessen's findings for discovery of  nuclides papers \cite{12Th}.  

A limited number of operating facilities combined with the constant levels of 
research staff and graduate students have led to more sophisticated experiments that require larger teams. These teams provide 
a fruitful environment for graduate students training, and increase in efficiency of the existing operations. Finally, 
the  extensive usage of  Web collaboration tools  made worldwide cooperation and data analysis possible, eventually leading to   
large geographically-separated groups of physicists working on a single project. That in turn increased the authorship of nuclear physics publications.

Other worth noticing nuclear physics trends: the largest number of articles per single scientist in NSR is 1017, and 100 out 
of 92600 authors have at least 330 publications. Such a large number of publications per single contributor could not be possible previously, when users operated a large number of small facilities. The overall number of nuclear reaction measurements in the area 1 is 37.16$\%$, area 2 is 34.79$\%$, 
area 3 is 13.35$\%$ and area 4 is 14.70$\%$ of the total. There is a recent increase of cross section measurements in areas 2 and 3. 
Hopefully, the LANSCE accelerator upgrade and full implementation of FRIB project will further increase an output for the area 1. 
The described above trends reflect a non-uniform distribution of nuclear science activities around the world. The present-day-situation in experimental nuclear 
physics authorship is similar to high energy physics, where large teams of physicists 
operate a small number of facilities \cite{14Aa}.

\section{Conclusions}
Results of the NSR and EXFOR bibliographical data mining have been presented. Large collections of bibliographical metadata  
represent a very powerful tool for understanding of the past, present, and, perhaps, future  research trends. The data analysis shows a 
strong anticorrelation between authorship increase of experimental papers and overall reduction of measurements due to closures of many small facilities.  
Presently, experimental groups often operate the sophisticated experiments at large user facilities.  The increase of the group sizes coincides in 
time with the development of Web collaboration tools.   These findings suggest that 
article authorship is a very complex phenomenon, and presently-observed increase or ``inflation" in authorship could be explained by  the adaptation to 
the changing research environment, in addition to the evolving authorship rules that progressed over the years  from very strict to lenient.  
An additional research  is necessary to investigate these new trends in other areas of science.

\section{Acknowledgments}                               
The author is grateful to U.S. Nuclear Data Program members, V. Zerkin (IAEA) and M. Blennau (BNL) for  productive discussions and  careful reading of 
the manuscript and useful suggestions, respectively. This work was funded by the Office of Nuclear Physics, Office of Science of 
the U.S. Department of Energy, under Contract No. DE-AC02-98CH10886 with Brookhaven Science Associates, LLC.  


\end{document}